%% file: main.tex
\documentclass[sigconf]{acmart}
\copyrightyear{2026}
\acmYear{2026}
\setcopyright{cc}
\setcctype{by}
\acmConference[EASE '26]{International Conference on Evaluation and Assessment in Software Engineering}{June 09--12, 2026}{Glasgow, United Kingdom}
\acmBooktitle{International Conference on Evaluation and Assessment in Software Engineering (EASE '26), June 09--12, 2026, Glasgow, United Kingdom}
\acmDOI{10.1145/3816483.3816556}
\acmISBN{979-8-4007-2348-3/2026/06}

\usepackage[utf8]{inputenc}
\usepackage[T1]{fontenc}
\usepackage[english]{babel}

\usepackage{colorblind}

\usepackage{booktabs}
\usepackage{multirow}
\usepackage{nicematrix}
\usepackage{tabularx}
\usepackage{cleveref}
\usepackage{siunitx}

\usepackage{tikz}
\usetikzlibrary{positioning, arrows, arrows.meta, decorations.pathreplacing, calc, backgrounds, fit, shapes, decorations.pathmorphing, fadings, shadings}

\usepackage{float}

\usepackage{pgfplots}
\pgfplotsset{compat=1.18}
\usepackage{pgfplotstable}

\newcommand{\textquote}[1]{\textit{``#1''}}

\newcommand{\numparticipants}{100}
\newcommand{\numcompaniestext}{five}

\input{data/results.tex}

\begin{document}

\title{Quo Vadis, Code Review? Exploring the Future of Code Review}

\author{Michael Dorner}
\affiliation{%
  \institution{Technische Hochschule Nürnberg}
  \city{Nürnberg}
  \country{Germany}
}
\email{michael.dorner@th-nuernberg.de}

\author{Andreas Bauer}
\affiliation{%
  \institution{Technische Hochschule Nürnberg}
  \city{Nürnberg}
  \country{Germany}
}
\email{andreas.bauer@th-nuernberg.de}

\author{Darja Šmite}
\affiliation{%
  \institution{Blekinge Institute of Technology}
  \city{Karlskrona}
  \country{Sweden}
}
\email{darja.smite@bth.se}

\author{Lukas Thode}
\affiliation{%
  \institution{Blekinge Institute of Technology}
  \city{Karlskrona}
  \country{Sweden}
}
\email{lukas.thode@bth.se}

\author{Daniel Mendez}
\affiliation{%
  \institution{Blekinge Institute of Technology}
  \city{Karlskrona}
  \country{Sweden}
}
\affiliation{%
  \institution{fortiss}
  \city{München}
  \country{Germany}
}
\email{daniel.mendez@bth.se}

\author{Ricardo Britto}
\affiliation{%
  \institution{Ericsson}
  \city{Stockholm}
  \country{Sweden}
}
\affiliation{%
  \institution{Blekinge Institute of Technology}
  \city{Karlskrona}
  \country{Sweden}
}
\email{ricardo.britto@bth.se}

\author{Stephan Lukasczyk}
\affiliation{%
  \institution{JetBrains Research}
  \city{München}
  \country{Germany}
}
\email{stephan.lukasczyk@jetbrains.com}

\author{Ehsan Zabardast}
\affiliation{%
  \institution{Blekinge Institute of Technology}
  \city{Karlskrona}
  \country{Sweden}
}
\email{ehsan.zabardast@bth.se}

\author{Michael Kormann}
\affiliation{%
  \institution{SAP}
  \city{München}
  \country{Germany}
}
\email{michael.kormann@sap.com}

\renewcommand{\shortauthors}{Dorner et al.}

\begin{abstract}
\noindent\emph{Context:}
Code review has long been a core practice in collaborative software engineering. As automation becomes increasingly embedded in development workflows, the role and functioning of code review are subject to change.

\noindent\emph{Objective:}
This study explores how professional developers anticipate the evolution of code review and identifies emerging tensions reflected in these expectations. 

\noindent\emph{Method:} 
We conducted a cross-sectional survey with \numparticipants{} developers across \numcompaniestext{} software-driven companies. The survey captured estimates of current review time and reviewed artifacts, as well as anticipated changes over a five-year horizon. Open-ended questions invited reflections on the future of code review. Quantitative responses were analyzed descriptively, and open-ended responses were independently coded by multiple researchers using thematic analysis to identify recurring patterns in participant responses.

\noindent\emph{Results:} 
Practitioners expect code review to remain essential, anticipating stable or increased time investment and a broader range of reviewed artifacts over the next five years. In open-ended responses, many participants explicitly referenced AI and large language models (LLMs), describing increasing automation in both code authoring and reviewing, including scenarios in which automated systems operate in both roles.

\noindent\emph{Conclusion:} 
Our analysis suggests emerging tensions concerning understanding, accountability, and trust in automation-mediated code review. These tensions provide early empirical signals of socio-technical challenges and position code review as a concrete setting for examining the implications of LLM integration in collaborative software engineering.
\end{abstract}



\acmCodeLink{https://github.com/michaeldorner/quo-vadis-code-review}
\acmDataLink{https://github.com/michaeldorner/quo-vadis-code-review}

\maketitle

\section{Introduction}

During the early days of software engineering, code was often designed and implemented by individual developers. As software systems became more complex, software engineering evolved into a collaborative effort involving multiple developers with diverse skills, organized in teams. To maintain a shared understanding of the codebase and its changes, developers discuss code changes before they are integrated.

As software engineering scaled from individual to collaborative and globally distributed work, these discussions evolved accordingly. In the 1970s, they were structured, in-person, and formal, and known as \emph{code inspections}~\cite{Fagan1976}. In the 1990s and 2000s, \emph{pair programming} integrated a synchronous form of those discussions between two developers directly into development workflows~\cite{Williams2001}. As development became increasingly distributed, discussions shifted toward asynchronous, tool-supported formats. Today, they are commonly referred to as \emph{code reviews}, a widely adopted practice in collaborative software engineering~\cite{Bauer2023,Dorner2024upperbound}.

While prior research has predominantly framed the practice as human-centered and collaborative \cite{Dorner2024upperbound,Bauer2023,Bacchelli2013}, automation in software engineering has expanded steadily. Recent advances in large language models (LLMs) have accelerated this trajectory. Empirical studies have begun evaluating the quality and usefulness of LLM-generated review comments~\cite{Alami2025}, and practitioner-oriented reports document active exploration of generative AI tools in code review contexts~\cite{Davila2024}. At the same time, LLMs are increasingly used for code generation~\cite{Barke2023}, raising the possibility that automation may extend into both authoring and reviewing. This shift raises new questions about how the role and configuration of code review may evolve.

The objective of this study is to explore how professional developers anticipate the evolution of code review and identify emerging tensions reflected in these expectations. 

To address this objective, we conducted a cross-sectional survey with \numparticipants{} across \numcompaniestext{} companies. As changes in code review cannot be directly observed within such a design, we operationalized them through participants’ reported current practices and anticipated changes in time spent on code review, reviewed artifacts, and expected changes to code review and their implications over the next five years.

This study contributes (1) empirical evidence on anticipated evolution of code review, (2) a conceptual model of human–AI role configurations, and (3) an analysis of emerging tensions in understanding, accountability, and trust.

The remainder of this paper presents the survey design and analysis approach (\Cref{sec:method}), reports empirical findings on anticipated evolution of code review (\Cref{sec:results}), analyzes emerging tensions arising from these findings (\Cref{sec:discussion}), and concludes with implications for research and practice (\Cref{sec:conclusion}).

\section{Method}
\label{sec:method}

We conducted an exploratory cross-sectional survey to examine how professional software developers perceive code review today and how they anticipate its evolution over the next five years.

\subsection{Sampling Strategy}

Because no comprehensive sampling frame of professional software developers exists~\cite{Baltes2022}, we employed \emph{quota sampling}, a non-random two-stage sampling strategy. First, we purposively selected \numcompaniestext{} companies representing diverse industry domains, primary software system types, and organizational scales. Second, we aimed to recruit approximately 25 developers per company. Recruitment concluded after eight weeks or once the target quota was reached. 

\subsection{Sample}

The final sample comprises \numparticipants{} professional software developers from \numcompaniestext{} companies, including SAP, Ericsson, JetBrains, Gradle, and one large European bank that requested anonymity (see \Cref{tab:company-background}). All respondents reported involvement in code review within their organizations. Participation varied by company due to voluntary recruitment. Responses are analyzed at the aggregate level, where companies serve as contextual background rather than analytical units.

\begin{table*}
  \renewcommand{\arraystretch}{1.4}
  \caption{Companies of the participating developers and sample sizes per company}
  \centering
  \begin{tabularx}{\linewidth}{@{}p{2.9cm}XXXXXXX@{}}
    \toprule
    \textbf{Attribute} 
      & \textbf{SAP} 
      & \textbf{Ericsson} 
      & \textbf{\textit{A Bank}} 
      & \textbf{JetBrains}
      & \textbf{Gradle} \\
    \midrule
    Domain
      & Enterprise software 
      & Telecommunications 
      & Banking and finance 
      & Developer Tools
      & Developer Tools \\
    Software System~Type
 
      & Customer-facing enterprise platforms 
      & Embedded and network control software 
      & Internal financial transaction and risk management systems 
      & Developer productivity and language tooling
      & Build automation and developer productivity tooling \\
    \# of employees 
      & >100{,}000 
      & >100{,}000 
      & >25{,}000 
      & >2{,}200
      & >150 \\
    \# of participants 
      & 31 
      & 24 
      & 25 
      & 11
      & 9 \\
    \bottomrule
  \end{tabularx}
  \label{tab:company-background}
\end{table*}

\subsection{Survey Instrument}

The survey captured both perceptions of current code review and anticipated changes over a five-year horizon. The five-year timeframe was selected to balance near-term plausibility with sufficient scope for structural change.

To capture participants' perceptions of current code review practices, we collected self-reported data on:
\begin{itemize}
    \item Average weekly time spent on code review (numerical response)
    \item Types of software artifacts currently reviewed (multiple selection)
\end{itemize}

To capture anticipated evolution, participants indicated:
\begin{itemize}
    \item Whether they expect to spend more, less, or the same amount of time on code review in five years (closed-end)
    \item Which artifacts they anticipate reviewing in five years (multiple selection).
    \item What major changes they anticipate in code review (open-ended)
    \item What implications they foresee based on those changes (open-ended)
\end{itemize}

Artifact categories included production code, test code, configuration files, documentation, and GUI-based test code. Participants could add additional artifact types.

\subsection{Data Collection}

The survey was conducted between December 2024 and November 2025.
For each company, we set up an independent instance of the online questionnaire and distributed it through the company's internal communication channels.
Recruitment was conducted sequentially across organizations, with each company-specific survey window remaining open for up to eight weeks or until the targeted quota was reached. Participation was voluntary and anonymous. To preserve confidentiality, we did not collect demographic information such as age, gender, or role seniority. Participation counts per company are reported without attributing responses to individuals or specific organizations.

\subsection{Data Analysis}

Quantitative responses were analyzed descriptively to summarize current time investment and reviewed artifacts, as well as anticipated changes over the five-year horizon.

Open-ended responses were analyzed using a reflexive form of thematic analysis, following the approach outlined by~\citeauthor{Clarke2017}~\cite{Clarke2017}. The first author conducted inductive descriptive coding. Initially, all responses were reviewed to ensure familiarity with the data. Preliminary codes were then assigned to participants' descriptions of anticipated changes and perceived implications. Codes were iteratively refined as additional responses were analyzed. 

Subsequently, codes were examined and grouped into broader themes based on conceptual or semantic similarities. A theme captures ``something important about the data in relation to the research question, and represents some level of patterned response or meaning within the data set''~\cite{braun2006UsingThematicAnalysis}.

The second author reviewed the codes to develop broader themes. Consistent with reflexive thematic analysis, we moved away from independent parallel coding or inter-rater reliability in favor of collaborative discussion. Throughout this process, we drew on our professional backgrounds in empirical software engineering to reflexively interpret the data through a socio-technical lens. Rather than reporting the numerical prevalence of themes, we concentrated on the conceptual importance across all responses to ensure that findings like accountability and trust were grounded in both participant reflections and professional context.

\section{Results}
\label{sec:results}

This section presents descriptive findings on (1) current weekly time spent on code review and expected changes over a five-year horizon, (2) reviewed artifacts today and anticipated future coverage, and (3) qualitative patterns emerging from open-ended responses regarding anticipated changes and their implications. All findings reflect self-reported perceptions and expectations rather than observed longitudinal developments.

\subsection{Effort and Scope of Code Review}

Across \numparticipants{} respondents, the median self-reported time spent on code review is approximately three hours per week (IQR: 2--6). Most respondents report spending fewer than five hours per week, while a smaller subset reports substantially higher time investment of up to 16 hours.

When asked how their review time may change over the next five years, 
\moretime{} respondents expect to spend \emph{more time}, 
\sametime{} expect to spend \emph{about the same time}, and 
\lesstime{} expect to spend \emph{less time} on code review. 

Respondents also anticipate a broader scope of reviewed artifacts. \Cref{fig:artifacts} shows paired percentages for artifacts reviewed \emph{today} versus those expected \emph{in five years}. Production code remains dominant (86\% today; 89\% expected), while increases are anticipated for test code, configuration files, documentation, and GUI-based test code. The share reporting that they review no artifacts decreases slightly.

\begin{figure}
  \begin{tikzpicture}
    \begin{axis}[
        axis line style={draw=none},
        width=\columnwidth,
        height=1.22*\columnwidth,
        xtick={0,1},
        xticklabels={Today,In Five Years},
        ymin=0, ymax=100,
        legend style={at={(1.05,1)}, anchor=north west},
        ymajorgrids=true,
        yticklabel style={
          /pgf/number format/.cd,
          fixed,
          precision=0,
        },
        yticklabel={\pgfmathprintnumber{\tick}\%},
        clip=false,
        enlargelimits=true,
      ]
		\addplot[mark=*, thick] coordinates {(0, 86.0) (1, 89.0)} coordinate [pos=0.5] (artifact0);
		\node[font=\small, above=1em] (artifactlabel0) at (artifact0) {Production code};
		\draw (artifact0) -- (artifactlabel0);
		
		\addplot[mark=*, thick] coordinates {(0, 64.0) (1, 70.0)} coordinate [pos=0.05] (artifact1);
		\node[font=\small, above=1em] (artifactlabel1) at (artifact1) {Test code};
		\draw (artifact1) -- (artifactlabel1);
		
		\addplot[mark=*, thick] coordinates {(0, 63.0) (1, 72.0)} coordinate [pos=0.95] (artifact2);
		\node[font=\small, above=1em, text width=2.5cm, align=center] (artifactlabel2) at (artifact2) {Parameter and configuration files};
		\draw (artifact2) -- (artifactlabel2);
		
		\addplot[mark=*, thick] coordinates {(0, 56.0) (1, 64.0)} coordinate [pos=0.5] (artifact3);
		\node[font=\small, below=1em] (artifactlabel3) at (artifact3) {Documentation};
		\draw (artifact3) -- (artifactlabel3);
		
		\addplot[mark=*, thick] coordinates {(0, 17.0) (1, 26.0)} coordinate [pos=0.5] (artifact4);
		\node[font=\small, above=1em] (artifactlabel4) at (artifact4) {GUI-based test code (end-to-end testing)};
		\draw (artifact4) -- (artifactlabel4);
		
		\addplot[mark=*, thick] coordinates {(0, 5.0) (1, 3.0)} coordinate [pos=0.5] (artifact5);
		\node[font=\small, above=1em] (artifactlabel5) at (artifact5) {None};
		\draw (artifact5) -- (artifactlabel5);
		
		\addplot[mark=*, thick] coordinates {(0, 1.0) (1, 1.0)} coordinate [pos=0.5] (artifact6);
		\node[font=\small, below=1em] (artifactlabel6) at (artifact6) {Others};
		\draw (artifact6) -- (artifactlabel6);
    \end{axis}
  \end{tikzpicture}

  \caption{Expected change in reviewed artifacts: respondents expect broader review coverage, with the largest increase for GUI-based test code. Percentages reflect multiple selection.}
  \label{fig:artifacts}
\end{figure}

Taken together, responses indicating stable or increasing review effort substantially outnumber those anticipating a decrease. Similarly, expectations of an expansion in reviewed artifacts exceed those anticipating contraction.

\subsection{AI as a Participant in Code Review}

In open-ended responses on the anticipated changes, many participants explicitly referenced AI and LLMs, even though the survey did not explicitly focus on AI. Rather than replacing human reviewers entirely, participants describe configurations in which AI performs early filtering or routine assistance before human inspection.

For example, one respondent noted that \textquote{AI automated review will become standard to identify potential pitfalls automatically -- probably before the human review happens.} Another suggested \textquote{Initial code review to be done by some form of an AI that can pick some basic stuff that are often missed by users.} Similarly, participants described minor issues being \textquote{sorted out with help of AI tools well before code review} and AI performing a \textquote{pre-review, e.g., highlighting the important points or adding explanations.}

In these accounts, AI is expected to handle syntactic checks, common defects, and routine issues, while human reviewers focus on architectural decisions, domain logic, and solution-level concerns. Several respondents explicitly anticipated a shift away from style-level discussions toward higher-level reasoning, with one stating that review would focus \textquote{more on architectural aspects instead of code style aspects.}

These responses point to a shift in the distribution of responsibilities within code review, where automation supports routine checks and human reviewers remain central to higher-level evaluation.

\subsection{More Code to Review}

Beyond the integration of AI into code review, many participants anticipate a substantial increase in the volume of generated code and pull requests due to AI-assisted coding. This increase is expected to raise the overall demand for code review.

A respondent observed that \textquote{AI will produce a tons of code with various quality. Reading that code will be more important than actually writing it.} Another anticipated \textquote{an overwhelming amount of reading that we'll have to do (less writing and tons more of reading).} Others explicitly connected AI-assisted coding with increased review workload, noting that \textquote{more code to review because of AI assisted coding} and \textquote{spending more time on code review rather than coding.}

These responses suggest that AI-assisted coding may not reduce code review effort, but instead increase the volume of code requiring reviews. Therefore, developers anticipate spending more time assessing and validating generated code. This expectation aligns with the quantitative finding that stable or increasing review time substantially outweighs expectations of decline.

\subsection{Fully Automated Configurations}

While many responses describe hybrid configurations, some participants explicitly reflected on the possibility of simultaneous AI authoring and reviewing. These reflections articulate a boundary scenario in which both code generation and code review are delegated to automated systems.

One participant cautioned: \textquote{We will have to decide if we want to write code with AI or review code with AI [...] I don't think both at the same time will work [...] the AI doesn't catch the mistake because it generated it.} Another described the prospect of \textquote{AI will be used to review code, and then we will have to review code and also the AI review of the code.}

Such responses reflect concerns about potential feedback loops and error propagation when automated systems operate in both authoring and reviewing roles. We conceptualize these role combinations along two intersecting continua: code author (human--LLM) and code reviewer (human--LLM). As illustrated in \Cref{fig:newactor}, this yields four possible role combinations, including a fully automated boundary case. We present this boundary case as a conceptual extreme derived from participant reflections rather than as a prediction of near-term practice.

\begin{figure}
  \centering
  \begin{tikzpicture}[
      x=0.85\columnwidth, y=0.85\columnwidth,
      cluster/.style={
        white,
        font=\bfseries\Large,
        text width=3.cm,
        text centered,
        inner sep=4mm
      }
    ]

    \coordinate (humanreviewer) at (0,1.1) ;
    \coordinate (llmreviewer) at (1,1.1) ;


    \draw[thick, stealth-stealth] (humanreviewer) -- (llmreviewer) node[midway, anchor=mid, fill=white] {Code Reviewer} ;

    \node[anchor=north west] at (humanreviewer) {Human};
    \node[anchor=north east] at (llmreviewer) {LLM};

    \coordinate (llmauthor) at (-0.1,0) ;
    \coordinate (humanauthor) at (-0.1,1) ;

    \draw[thick, stealth-stealth] (humanauthor) -- (llmauthor) node[midway, rotate=90, anchor=mid, fill=white] {Code Author} ;

    \node[anchor=north east, rotate=90] at (humanauthor) {Human};
    \node[anchor=north west, rotate=90] at (llmauthor) {LLM};

    \shade [
      upper left=blue,
      upper right=green,
      lower left=cyan,
      lower right=yellow,
    ] (0,0) rectangle (1,1);

    \draw (0,0) rectangle (1,1);

    \node[cluster, anchor=north west] at (0, 1) {Human-led\\software engineering};
    \node[cluster, anchor=north east] at (1, 1) {Automated code review};
    \node[cluster, anchor=south west] at (0, 0) {Automated code generation};
    \node[cluster, anchor=south east] at (1, 0) {Unsupervised software engineering};
  \end{tikzpicture}
  \caption{Code review conceptualized along two continuous dimensions: author and reviewer roles ranging from fully human-led to fully LLM-led. The resulting quadrants represent alternative configurations of human–AI collaboration suggested by participant responses.}
  \label{fig:newactor}
\end{figure}

\subsection{Human Oversight and Responsibility}

Despite expectations of increasing automation, multiple participants emphasized that human review will remain essential. Respondents did not describe a future without code review, but one in which AI participation coexists with continued human responsibility.

For example, one participant stated, \textquote{I do not think that code review will disappear completely due to agents or LLMs.} Another argued that \textquote{human involvement will be much more important,} particularly when AI-generated code is involved. Several responses also referenced legal and organizational accountability, suggesting that AI-generated code would still require human review to ensure responsibility, including to \textquote{avoid any [lawsuits] on the code development organization.}

These responses indicate that, even in settings with substantial automation, practitioners expect human oversight to remain central to code review. While AI may assist with routine checks or preliminary analysis, responsibility for evaluating correctness, domain relevance, and organizational risk is expected to remain with human reviewers.

\section{Discussion}
\label{sec:discussion}

Our findings indicate that practitioners do not expect code review to decline. Instead, they anticipate stable or increasing time investment, broader artifact coverage, and a growing volume of AI-generated code requiring review. At the same time, participants expect increasing integration of AI, typically in supportive roles rather than as full replacements for human reviewers. We interpret those insights as early empirical signals of change in code review and discuss their implications along three dimensions of code review: understanding, accountability, and trust.

\subsection{Understanding}

Several respondents expressed concern that increasing reliance on AI-generated code may alter how developers understand the systems they maintain. Expectations of reviewing larger volumes of generated artifacts and spending more time reading than writing suggest a shift in how developers engage with code. 

Code review has traditionally served not only as a quality assurance mechanism, but also as a communication network and collaborative practice~\cite{Dorner2024upperbound,Fatima2019,Mcintosh2016}. In layered configurations where AI performs preliminary checks and developers focus on higher-level concerns, the nature of understanding may shift from line-by-line scrutiny toward validation of generated solutions. This shift does not necessarily imply a decline in comprehension. However, it may change how, where, and potentially even whether the same depth of understanding is developed within human teams.

Concerns about diminishing experiential learning were also visible in responses suggesting that AI-assisted generation may reduce opportunities for developers to engage deeply with implementation details. Such expectations resonate with recent discussions of AI-generated code as a form of generative reuse~\cite{Taivalsaari2025}, in which developers increasingly integrate artifacts they did not design. If the anticipated scale increase materializes, maintaining long-term human comprehension may require deliberate practices to keep pace with the growing volume of generated artifacts.

\subsection{Accountability}

Despite expectations of increasing automation, respondents repeatedly emphasized the continued necessity of human oversight. Some explicitly referred to legal exposure and organizational responsibility, suggesting that AI-generated code would remain subject to human review to ensure accountability.

Code review has long functioned as a mechanism for distributing and reinforcing responsibility within teams~\cite{Bacchelli2013,Zabardast2022}. In layered human--AI configurations, however, evaluative work may be partially delegated to tools that cannot themselves bear responsibility. While human developers remain formally accountable, the evaluative authority may become less clearly defined when AI generates code, highlights defects, or proposes fixes.

Rather than eliminating accountability, automation may redistribute it. Human reviewers may increasingly evaluate not only code changes, but also AI-generated suggestions and analyses. This dual evaluation structure, reflected in responses describing the need to \textquote{review the AI review}, suggests a potential expansion of oversight responsibilities. As regulatory frameworks increasingly emphasize traceability and governance in software engineering~\cite{Dorner2024tax,Sauvola2024}, ensuring transparent lines of human responsibility in AI-augmented code review may become increasingly critical.

\subsection{Trust}

Trust emerged implicitly in several responses reflecting skepticism toward fully automated configurations. Participants questioned the simultaneous delegation of authoring and reviewing to AI, highlighting concerns about feedback loops and error propagation. The scenario in which AI-generated code is subsequently reviewed by another AI was described as potentially problematic, particularly if errors introduced during generation remain undetected.

Trust in code review traditionally rests on interpersonal knowledge, shared context, and identifiable contributors. In AI-augmented settings, distinguishing between human judgment and automated suggestion may become less straightforward. As one participant noted, the involvement of AI may require additional scrutiny rather than less, potentially making review more complex rather than simpler.

Empirical studies have already shown that developers assess AI-generated review comments differently from human feedback~\cite{Alami2025human}. Our findings suggest that such distinctions may become increasingly relevant as AI participation expands. Maintaining calibrated trust in hybrid human--AI review processes may therefore require clearer attribution of contributions, explicit norms governing AI use, and transparency about tool involvement.

\section{Threats to Validity}
\label{sec:threats}

As an exploratory mixed-methods study, our research design involves inherent trade-offs. A primary limitation is our reliance on self-reported perceptions of the future. To mitigate the highly speculative nature of this data, we operationalized anticipated evolution through concrete, comparative assessments, such as current versus anticipated time spent and artifacts reviewed, rather than asking for open-ended predictions of the distant future. 

Consequently, a key limitation is that our findings reflect practitioners’ expectations rather than observed changes. While such perceptions offer early signals, they are inherently speculative and may be influenced by the recent prominence of generative AI. Therefore, results should be interpreted as indicative of perceived trajectories, not validated developments.

Another notable threat stems from our sequential data collection across the \numcompaniestext{} participating companies from December 2024 to November 2025. While this extended period was a logistical necessity to accommodate organizational communication policies, it coincided with rapid advancements in generative AI tools. However, our analysis revealed consistent thematic patterns regarding core tensions (understanding, accountability, and trust) across early and late respondents. This suggests that practitioners' long-term expectations are anchored in fundamental collaborative practices rather than short-term fluctuations in tool capabilities. 

Furthermore, because no comprehensive sampling frame of software developers exists, we employed non-random quota sampling. While our sample of \numparticipants{} developers is not statistically representative of the global software engineering population, our purposive selection of companies representing distinct industry domains and varying scales ensures a diverse baseline of practitioner perspectives, supporting the transferability of our findings to similar environments.

Finally, our qualitative findings rely on the thematic analysis of open-ended responses and are inherently subject to researcher subjectivity. Consistent with reflexive thematic analysis, we did not attempt to eliminate this subjectivity via independent parallel coding. Instead, we leveraged our backgrounds in empirical software engineering to collaboratively and reflexively interpret the data, ensuring themes were grounded directly in participant accounts rather than preconceived notions about automation.

\section{Conclusion}
\label{sec:conclusion}

Code review has long been a cornerstone of collaborative software engineering, where experience, reasoning, and feedback shape the evolution of software systems. Our results indicate that it is unlikely to disappear in the foreseeable future. On the contrary, participants expect stable or increasing review effort and an expansion in the range of reviewed artifacts.

While code review will remain essential, it is expected to evolve substantially. In particular, participants anticipate that LLMs will become ubiquitous in development workflows and increasingly take on tasks in both code authoring and review. This raises the question to what extent evaluative responsibilities should be delegated to AI or remain with human developers. Rather than assuming purely beneficial effects, our findings point to emerging concerns regarding over-reliance on LLMs, including potential impacts on understanding, accountability, and trust.

As software engineering continues to evolve toward greater automation, we therefore advocate for a nuanced perspective on code review. While automation may improve efficiency, maintaining human understanding, accountability, and calibrated trust remains critical for sustaining control over the long-term evolution of software systems. These qualities, although less easily measurable or optimizable, underpin the resilience of software engineering in the face of ongoing technological change.

We also encourage further research into the implications of integrating LLMs and other generative AI techniques into code review, as this practice may serve as an early indicator of how AI reshapes collaborative software engineering more broadly. An over-reliance on AI may risk shifting the locus of understanding  from humans to the LLM---an entity that, despite its capabilities, cannot be held accountable, reason contextually, or justify its decisions beyond probabilistic associations. To preserve the integrity of software engineering as a collaborative, transparent, trustworthy, and responsible engineering practice, it remains essential that code review remains grounded in human understanding and oversight.

\section*{Data and Code Availability} 

The survey instrument, all anonymized survey responses and analysis scripts, as well as coding materials, are publicly available at: 
\url{https://github.com/michaeldorner/quo-vadis-code-review}

\balance

\begin{acks}
We thank all participants of the survey and the anonymous reviewers for their insightful and constructive feedback.
This work was supported by the KKS Foundation through the SERT~Project (Research Profile Grant 2018/010) at Blekinge Institute of Technology.	
\end{acks}

\bibliographystyle{ACM-Reference-Format}
\bibliography{references.bib}

\end{document}

%% file: data/results.tex
\newcommand{\moretime}{\SI{47}{\percent}}
\newcommand{\sametime}{\SI{30}{\percent}}
\newcommand{\lesstime}{\SI{23}{\percent}}

%% file: references.bib
@inproceedings{Bacchelli2013,
	title = {Expectations, outcomes, and challenges of modern code review},
	author = {Bacchelli, Alberto and Bird, Christian},
	year = 2013,
	booktitle = {2013 35th International Conference on Software Engineering (ICSE)},
	volume = {},
	number = {},
	pages = {712--721},
	doi = {10.1109/ICSE.2013.6606617}
}

@article{Baltes2022,
	title = {Sampling in software engineering research: a critical review and guidelines},
	author = {Sebastian Baltes and Paul Ralph},
	year = 2022,
	month = 7,
	journal = {Empirical Software Engineering},
	publisher = {Springer},
	volume = 27,
	pages = 94,
	doi = {10.1007/s10664-021-10072-8},
	issn = {1382-3256},
	issue = 4
}

@article{Bauer2023,
	title = {Code review guidelines for GUI-based testing artifacts},
	author = {Andreas Bauer and Riccardo Coppola and Emil Alégroth and Tony Gorschek},
	year = 2023,
	month = 11,
	journal = {Information and Software Technology},
	volume = 163,
	pages = 107299,
	doi = {10.1016/j.infsof.2023.107299},
	issn = {09505849}
}

@article{Clarke2017,
	title = {Thematic analysis},
	author = {Victoria Clarke and Virginia Braun},
	year = 2017,
	month = 5,
	journal = {The Journal of Positive Psychology},
	volume = 12,
	pages = {297--298},
	doi = {10.1080/17439760.2016.1262613},
	issn = {1743-9760},
	issue = 3
}

@article{Davila2024,
	title = {Tales From the Trenches: Expectations and Challenges From Practice for Code Review in the Generative AI Era},
	author = {Nicole Davila and Jorge Melegati and Igor Wiese},
	year = 2024,
	month = 11,
	journal = {IEEE Software},
	volume = 41,
	pages = {38--45},
	doi = {10.1109/MS.2024.3428439},
	issn = {0740-7459},
	issue = 6
}

@article{Dorner2024tax,
	title = {Taxing Collaborative Software Engineering},
	author = {Michael Dorner and Maximilian Capraro and Oliver Treidler and Tom-Eric Kunz and Darja Šmite and Ehsan Zabardast and Daniel Mendez and Krzysztof Wnuk},
	year = 2024,
	journal = {IEEE Software},
	pages = {1--8},
	doi = {10.1109/MS.2023.3346646},
	issn = {0740-7459}
}

@article{Dorner2024upperbound,
	title = {The Upper Bound of Information Diffusion in Code Review},
	author = {Michael Dorner and Daniel Mendez and Krzysztof Wnuk and Ehsan Zabardast and Jacek Czerwonka},
	year = 2023,
	month = 6,
	journal = {Empirical Software Engineering},
	publisher = {Springer}
}

@article{Fagan1976,
	title = {Design and code inspections to reduce errors in program development},
	author = {M. E. Fagan},
	year = 1976,
	journal = {IBM Systems Journal},
	volume = 15,
	pages = {182--211},
	doi = {10.1147/sj.153.0182},
	isbn = 3540430814,
	issn = {0018-8670},
	issue = 3,
	pmid = 11441438
}

@article{Sauvola2024,
	title = {Future of software development with generative AI},
	author = {Sauvola, Jaakko and Tarkoma, Sasu and Klemettinen, Mika and Riekki, Jukka and Doermann, David},
	year = 2024,
	journal = {Automated Software Engineering},
	volume = 31,
	number = 1,
	doi = {10.1007/s10515-024-00426-z},
}

@inproceedings{Williams2001,
	title = {Integrating pair programming into a software development process},
	author = {Laurie Williams},
	booktitle = {Proceedings 14th Conference on Software Engineering Education and Training. 'In search of a software engineering profession' (Cat. No.PR01059)},
	publisher = {IEEE Comput. Soc},
	pages = {27--36},
	doi = {10.1109/CSEE.2001.913816},
	isbn = {0-7695-1059-0}
}

@inproceedings{Zabardast2022,
	title = {Ownership vs Contribution: Investigating the Alignment Between Ownership and Contribution},
	author = {Ehsan Zabardast and Javier Gonzalez-Huerta and Binish Tanveer},
	year = 2022,
	month = 3,
	booktitle = {2022 IEEE 19th International Conference on Software Architecture Companion (ICSA-C)},
	publisher = {IEEE},
	pages = {30--34},
	doi = {10.1109/ICSA-C54293.2022.00013},
	isbn = {978-1-6654-9493-9},
}

@inproceedings{Fatima2019,
	title = {Knowledge Sharing, a Key Sustainable Practice Is on Risk: {{An}} Insight from {{Modern Code Review}}},
	shorttitle = {Knowledge Sharing, a Key Sustainable Practice Is on Risk},
	author = {Fatima, Nargis and Nazir, Sumaira and Chuprat, Suriayati},
	year = 2019,
	month = dec,
	booktitle = {2019 {{IEEE}} 6th {{International Conference}} on {{Engineering Technologies}} and {{Applied Sciences}} ({{ICETAS}})},
	publisher = {IEEE},
	address = {Kuala Lumpur, Malaysia},
	pages = {1--6},
	doi = {10.1109/ICETAS48360.2019.9117444},
	isbn = {978-1-7281-4082-7},
}

@article{Mcintosh2016,
	title = {An Empirical Study of the Impact of Modern Code Review Practices on Software Quality},
	author = {McIntosh, Shane and Kamei, Yasutaka and Adams, Bram and Hassan, Ahmed E.},
	year = 2016,
	month = oct,
	journal = {Empirical Software Engineering},
	volume = 21,
	number = 5,
	pages = {2146--2189},
	doi = {10.1007/s10664-015-9381-9},
	issn = {1382-3256, 1573-7616},
}

@article{Taivalsaari2025,
	title = {On the Future of Software Reuse in the Era of AI Native Software Engineering},
	author = {Taivalsaari, Antero and Mikkonen, Tommi and Pautasso, Cesare},
	year = 2025,
	journal = {arXiv preprint arXiv:2508.19834}
}

@inproceedings{Alami2025human,
	title = {Human and Machine: How Software Engineers Perceive and Engage with AI-Assisted Code Reviews Compared to Their Peers},
	author = {Alami, Adam and Ernst, Neil},
	year = 2025,
	booktitle = {2025 IEEE/ACM 18th International Conference on Cooperative and Human Aspects of Software Engineering (CHASE)},
	pages = {63--74},
	organization = {IEEE}
}

@article{Barke2023,
   author = {Shraddha Barke and Michael B. James and Nadia Polikarpova},
   doi = {10.1145/3586030},
   issn = {2475-1421},
   issue = {OOPSLA1},
   journal = {Proceedings of the ACM on Programming Languages},
   month = {4},
   pages = {85-111},
   publisher = {Association for Computing Machinery},
   title = {Grounded Copilot: How Programmers Interact with Code-Generating Models},
   volume = {7},
   year = {2023}
}

@article{Alami2025,
   author = {Adam Alami and Nathan Cassee and Thiago Rocha Silva and Elda Paja and Neil A. Ernst},
   month = {12},
   title = {Engagement in Code Review: Emotional, Behavioral, and Cognitive Dimensions in Peer vs. LLM Interactions},
   journal = {arXiv preprint arXiv:2512.05309},
   year = {2025}
}

@article{braun2006UsingThematicAnalysis,
  title = {Using Thematic Analysis in Psychology},
  author = {Braun, Virginia and Clarke, Victoria},
  year = {2006},
  month = jan,
  journal = {Qualitative Research in Psychology},
  volume = {3},
  number = {2},
  pages = {77--101},
  issn = {1478-0887, 1478-0895},
  doi = {10.1191/1478088706qp063oa},
  urldate = {2023-03-01},
  langid = {english}
}
